\documentstyle[prl,aps,epsfig,preprint]{revtex}

\draft

\newlength{\textwidthm}
\begin{document}
\title{Storage of light in atomic vapor}
\author{D.~F.~Phillips, A.~Fleischhauer, A.~Mair, R.~L.~Walsworth}
\address{Harvard-Smithsonian Center for Astrophysics,
                Cambridge, MA~~02138 }
\author{ M.~D.~Lukin}
\address{ITAMP, Harvard-Smithsonian Center for Astrophysics,
                Cambridge, MA~~02138 }

\date{\today}
\maketitle
\begin{abstract}
We report an experiment in which a light
pulse is decelerated and  trapped in a vapor of Rb atoms, stored
for a controlled
period of time, and then released on demand.
We accomplish this storage of light by dynamically reducing the group
velocity of
the light pulse to zero, so that the coherent excitation of the light is 
reversibly mapped into a collective Zeeman (spin) coherence of the Rb vapor.
\end{abstract}


\nobreak

Photons are the fastest and most robust carriers
of information, but they are difficult to localize and store.
The present Letter reports a proof-of-principle demonstration of a
technique \cite{stop,pol} to trap, store, and release
excitations carried by light pulses. Specifically, a pulse of
light which is several kilometers long in free space is compressed
to a length of a few centimeters and then converted into collective spin
excitations in a vapor of Rb atoms.  After a controllable storage time, the
process
is reversed and the atomic coherence is converted back into a light pulse.

The light-storage technique is based on the recently demonstrated phenomenon of
ultra-slow light group velocity \cite{slow}, which is made possible by
Electromagnetically
Induced Transparency (EIT) \cite{rev}.
In ``slow-light'' experiments an external optical
field (the ``control field'') is used to make an otherwise opaque medium become
transparent near an atomic resonance. A weak optical field (the ``signal
field'')
at a particular frequency and polarization can then propagate without
dissipation
and loss but with a substantially reduced group velocity. Associated with
slow light is a considerable spatial compression, which allows
a signal pulse to be almost completely localized in the atomic medium. In
addition, as the signal light
propagates the atoms are driven into a collective coherent superposition
of (typically) Zeeman or hyperfine states that is strongly coupled to the
light via a Raman transition.  The coupled
light and atomic excitations can be efficiently described as a single form
of dressed-state excitation known as the
``dark-state polariton'' \cite{pol}. In order to store a light pulse, we
smoothly turn off the control field, which
causes the dark-state polariton to be adiabatically  converted into a
purely atomic excitation (a collective Rb Zeeman
coherence in the experiment reported here) which is confined to the vapor
cell.  Turning the control field back on
reverses the process: the dark-state polariton is adiabatically restored to
an optical excitation, which can
then leave the sample cell.

A key feature of the light-storage method is its
non-destructive nature \cite{dis}. Specifically, the collective atomic spin
excitations do not couple to  electronic excited states and are thus immune to
spontaneous emission. Therefore, in the ideal limit, the light-storage process
is completely coherent, i.e., it is
expected that the
phase and quantum state of the signal pulse can be preserved. In practice
the storage
time is limited by the atomic coherence lifetime. In the present
experiment we were able to measure storage times up to $\sim$ 0.5 ms.
The non-destructive nature of the light-storage technique makes it
an attractive candidate for potential
applications involving coherent communication between distant
quantum-mechanical systems \cite{stop}.  

Before proceeding we note that
adiabatic passage has also been considered
for photon state storage in {\it single} atoms in the  context of cavity
QED \cite{QED}, and that encouraging experimental progress in this
direction has recently been
reported \cite{rempe}. In contrast, the light-storage technique demonstrated
here involves collective atomic excitations in an optically-dense, 
{\it many-atom}  system which is
considerably
more robust and reliable than the single-atom approach and is thus much
easier to implement experimentally. Efficient protocols
for quantum
computation \cite{block}, teleportation \cite{tele}, and squeezing
\cite{squ} using such excitations are
currently being actively investigated.  Other related work includes studies
of the ``freezing'' of light
in moving media \cite{fre} and investigations of efficient nonlinear
optical effects using atomic coherences
\cite{nlo}.

The present experiment can be understood qualitatively by considering a
``lambda'' configuration of three atomic states coupled by a pair of
optical
fields  (see Fig. 1a). Here the control field (Rabi-frequency $\Omega_c$)
and signal field ($\Omega_s$) are
represented, respectively, by right and left circularly polarized light
($\sigma_+$ and $\sigma_-$) derived from a
single laser beam.  These light fields couple pairs of Zeeman sublevels of
electronic ground state ($5^{2}S_{1/2}$) Rb
atoms  ($|-\rangle, |+\rangle$), with magnetic quantum numbers differing by
two, via the excited $5^{2}P_{1/2}$ state.   In this configuration the 
effective
two-photon detuning can be controlled
by applying an external magnetic field, which causes Zeeman level shifts
between $|-\rangle$ and $|+\rangle$.
The two optical fields drive the atoms into a superposition of Zeeman
sublevels (a ``dark state'') which at zero magnetic field is
decoupled from the light.

We performed light-storage experiments in atomic Rb vapor at
temperatures $\sim$ 70-90 $^o$C, which corresponds to atomic densities
$\sim 10^{11}-10^{12}$ cm$^{-3}$.
Under these conditions the 4 cm-long sample cell was normally completely
opaque for a
weak optical field
near the Rb $D_1$ resonance ($\simeq$ 795 nm).
We derived the control
and signal beams from the output of  an extended
cavity diode laser by carefully controlling the light polarization as shown
in the experimental schematic in
Fig. 1c.
For the data presented here we employed the $5^{2}S_{1/2}, F=2 \rightarrow
5^{2}P_{1/2},
F=1$ transition in $^{87}$Rb.  The control field was always much stronger
than the signal
field ($\Omega_c \gg \Omega_s$); hence most of the relevant atoms were
in the $5^{2}S_{1/2}, F=2, M_F = +2$ magnetic sublevel. In this case
the states $|-\rangle, |+\rangle$ of  the simplified 3-level model
correspond, respectively, to $|F=2,M_F = 0\rangle$
and $|F=2,M_F = +2\rangle$. By using a fast Pockels cell we slightly
rotated the polarization of
the input light to create
a weak pulse of left circularly polarized ($\sigma_-$) light, which served
as the signal field. We monitored transmission of
the $\sigma_-$ light pulse using a $\lambda/4$ waveplate and
polarizing beam splitter. In order to ensure long lifetimes of the atomic
Zeeman
coherences, we magnetically shielded the Rb cell and filled it with about 5
torr of He buffer gas.
We used a precision solenoid to control the static magnetic
field along the propagation direction of the optical beam.

We first consider the case of cw signal and control fields.
Fig. 1b displays a typical transmission spectrum for the signal
($\sigma_-$) field obtained by scanning the magnetic
field and thereby changing the effective two-photon detuning.
Note that due to the induced transparency the signal field transmission is
maximal
for zero magnetic field, even though most of atoms are in the
state $|+\rangle$. Outside of the transparency window  (magnetic fields
$>$ 20 mG) the Rb vapor is completely opaque to $\sigma_-$ light.

We next present a demonstration of light storage. Typical input $\sigma_-$
signal pulses
had a temporal length of $\sim$ 10 to 30 $\mu$s, corresponding to a
spatial length of several kilometers in free space. Upon entrance into the Rb
cell the signal pulse was spatially compressed by more than
five orders of magnitude, due to the reduction
in group velocity, as estimated from the observed change in pulse
propogation. In order to
trap, store, and release the signal pulse, we used an acousto-optic
modulator to turn off the control field
smoothly over about 3 $\mu$s while much of the signal pulse was contained
in the Rb cell.
After some
time interval, we turned the control field on again, thereby releasing the
stored portion of
the signal pulse.

An example of the observed light storage is shown in Fig. 2. Typically, two
time-resolved $\sigma_-$ signal pulses
were  registered by the photodetector
in the process of light storage and release.
First, a fraction of the signal pulse left the cell before
the control field was turned off, which resulted in an observed signal that
was not
affected by the storage operation (peak I in each plot of Fig. 2). This
untrapped light was delayed by about 30 $\mu$s
as compared to free-space propogation due to the slow group velocity
($v_g \sim 1 $ km/s). The second observed signal pulse was light 
that was stored in atomic
excitations for a time interval $\tau$.
Note that no output signal was observed as long as the control
field was off; rather, the released signal pulse was detected only after the
control field was
turned back on (peak II in each plot of Fig. 2). The controlled light
storage is the principal result of this
Letter.  We observed that the  amplitude of
the released signal pulse decreased as the $\tau$ increased. We could
resolve released light pulses
without signal averaging for storage intervals up to $\tau \simeq$ 0.5 ms.

We turn now to a theoretical interpretation of the experimental results.
We consider
the propagation of a signal pulse in an EIT medium (along the $z$ direction)
subject to a time-dependent
control field. We assume that the signal field is always weaker than
control field and that the the signal field group velocity is
always much smaller than the control field group velocity.
(The latter assumption allows us to neglect the retardation and spatial
dependence of the
control field.)

As noted above, the dynamical trapping of signal pulses can be understood
in terms
of dark-state polaritons. These are coupled superpositions of photonic and
spin wave-like excitations,
defined by a transformation:
\begin{eqnarray}
&&\Psi(z,t)=\cos\theta(t)\, \Omega_s(z,t) - \sin\theta(t)\, \sqrt{\kappa}\,
\rho_{-+}(z,t), \\
&&\cos\theta(t) = {\Omega_c(t) \over \sqrt{\Omega_c^2(t) + \kappa}}, \;
\sin\theta(t) = {\sqrt{\kappa} \over \sqrt{\Omega_c^2(t) + \kappa}}.
\nonumber
\end{eqnarray}
Here $\rho_{-+}$ is the atomic coherence between states $|-\rangle$ and
$|+\rangle$.  Also, $\kappa = 3n \lambda^2
\gamma_r c/8\pi$, where $n$ is the $^{87}$Rb density, $\lambda$ is the
wavelength and $\gamma_r$ is the natural
linewidth of the
$D_1$ transition, and $c$ is the free-space speed of light.

In the ideal limit, corresponding to vanishing
dephasing of the atomic coherence and perfect adiabatic following,
the polariton propagation is described by \cite{pol}
\begin{equation}
\Psi(z,t)=\Psi\left[z- \int^t_{t_0}\!\!\!{\rm d}t'
v_g(t'),t=t_0\right],
\label{sol}
\end{equation}
where the time-dependent group velocity is
\begin{equation}
v_g(t) = \cos^2(\theta) = {\Omega_c^2(t) \over \Omega_c^2(t) + \kappa}.
\label{group}
\end{equation}
In the stationary case, $v_g(t) = v_g^0$, and
after initial pulse compression at the entrance of the cell, the
polariton describes the well-known EIT-like propagation of coupled light and
atomic coherence.  Remarkably, however, when the
intensity of the control field is changed during the pulse's propagation
through the atomic medium, the polariton
can preserve its shape, amplitude, and  spatial length, while its
group velocity and the ratio of the light and matter components are altered.
In particular, when the group velocity is reduced to zero by turning off the
control beam, the polariton becomes purely atomic ($\cos\theta = 0$) and
its propogation is stopped. The state of the input light pulse is thereby
mapped into the atomic coherence
$\rho_{-+}$.  The coherence stored
within the cell $(0<z<L_{cell})$ after switching off the control field over
the time interval [$t_0,t_1$] is given by:
\begin{equation}
\rho_{-+}(z,t_1) = - \sqrt{{c \over v_g^0 \kappa}}
\Omega_s(z=0,{\int_{t_0}^{t_1} v_g(t') d t' -z \over v_g^0}).
\label{coh}
\end{equation}
If the control beam is turned back on after a storage interval
$\tau$, the polariton is accelerated and the atomic coherence $\rho_{-+}$
is mapped back into
light.  The released light pulse has a shape, amplitude, and spatial length
proportional
to the coherence {\it after} the storage interval:
\begin{equation}
\Omega_s(z,t) = - \cos\theta(t) \sqrt{\kappa}
\rho_{-+}(L_{cell}-\int_{t_2}^t v_g(t') d t',t_2),
\label{fi}
\end{equation}
where $t_2 = t_1 + \tau$.  In principle, complete storage and retrieval of
the input light pulse
is possible. To make a detailed comparison with experimental results, however,
deviations from this ideal limit must be considered: e.g., decay of the
Zeeman coherence and non-adiabatic
corrections.  Results of theoretical calculations for the conditions of our
experiment (Fig. 3) are in good
agreement with the measurements displayed in Fig. 2.

Under realistic
conditions, the light storage time is always limited by
loss of atomic coherence. In the present experiment, for example, the Rb atoms
diffuse through the buffer gas and escape from
the region of interaction with the light beam, leading to a coherence time
of $\sim$ 150 $\mu$s. Likewise, at high atom
densities, coherences decay due to spin-exchange collisions.

It is also important to consider the assumption of adiabatic following and
its effect on
the {\it dynamic} method for group velocity reduction, as compared to the
conventional, stationary approach
based on EIT.   Naively, Eq. (\ref{group}) indicates that long light pulse
delays can be
obtained by simply using a cw control field of sufficiently low intensity.  
However,
for pulses similar to those used in the dynamic trapping method
we failed to observe long delays ($\sim$ 100 $\mu$s) using the stationary
EIT technique. This failure is due to  {\it the
breakdown of adiabatic following}.

The essence of adiabatic following is that the light pulse
spectrum
should be contained within a relatively narrow transparency window
$\Delta \omega$ (Fig. 1b) to avoid loss and dissipation. The magnitude
of $\Delta \omega$ is determined by both the control field intensity
and the opacity of the atomic medium \cite{lukin97}. In conventional
EIT propagation, a weaker
control field induces a narrower transmission spectrum. For a fixed
bandwidth of the propagating
signal pulse, such spectral  narrowing causes absorption of certain pulse
spectral components
and inevitably destroys propagating light. This loss is consistent with our
experimental observations using
the stationary EIT technique
and is confirmed by our calculations (e.g., the dotted line in Fig. 3c),
which include non-adiabatic corrections
corresponding  to a finite width of the EIT spectrum.

The observed light pulses stored and released by {\it dynamic} reduction of
the group velocity
are obviously not destroyed in spite of the narrowing of the 
transparency window.
This important result is in agreement with the theoretical predictions of
Ref.\cite{pol}, where it
was pointed out that adiabatic following occurs as long as the product of the
propagation distance and the normal opacity of the medium
(i.e., the absorption length $l_{abs} = \kappa/(\gamma_{opt} c)$ where
$\gamma_{opt}$ is the total linewidth of the
optical transition) is smaller than the square of the spatial light pulse
length in the medium. In other words,
adiabaticity can be preserved with the dynamic light-storage method as long
as the input pulse bandwidth is within
the {\it initial} transparency window.
A remarkable feature of the dark-state polariton is
that its spatial length remains unchanged in the process of deceleration.
Hence, a dynamic reduction in group velocity is accompanied
by a {\it narrowing} of the polariton frequency spectrum (bandwidth).
In this case, adiabatic following occurs
even when the group velocity is reduced to zero, which is
in good agreement with the experimental results presented here.

In conclusion we have demonstrated that it is possible to control
the propagation
of light pulses in optically thick media by dynamically changing the group
velocity. In particular,
a light pulse can be trapped and stored in an atomic coherence; after a
controllable delay this coherence can be
converted back into a light pulse.

It is a pleasure to thank M.~Fleischhauer and S.~Yelin for many fruitful
ideas and collaboration on theoretical aspects of this work.
We also thank S.~Harris and M.O.~Scully for many stimulating
discussions. This work was partially supported by the
National Science Foundation and the Office of Naval Research.

After completion of this work we were informed by Prof. Hau that
light storage has also been observed in
ultra-cold atomic gas \cite{hau}.


\def\etal{\textit{et al.}}


\begin{figure}[ht]
\centerline{\epsfig{file=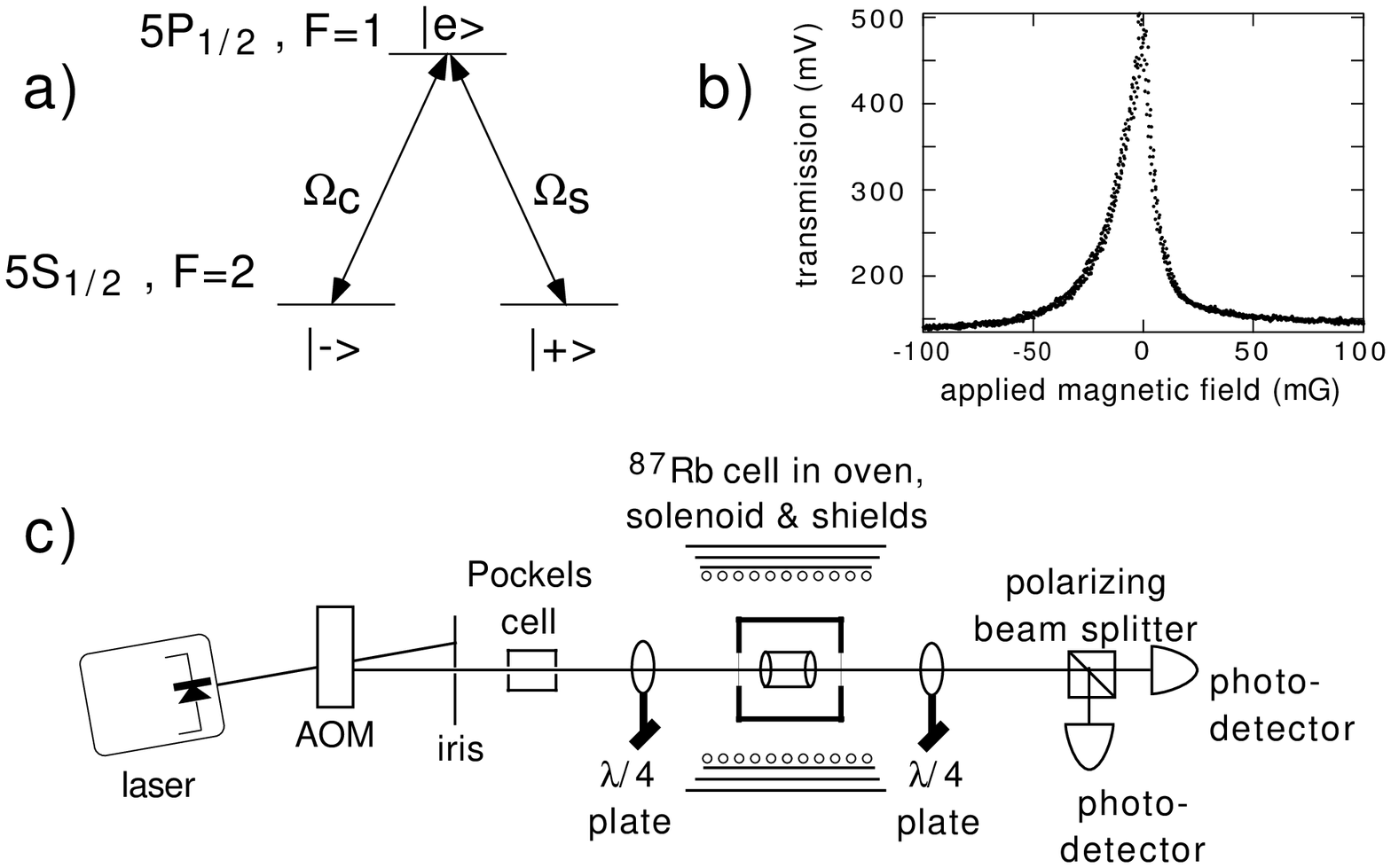,width=9.0cm}}
\vspace*{2ex}
\caption{(a) $\Lambda$-type configuration of $^{87}$Rb atomic states
resonantly coupled to a control field
  ($\Omega_c$) and a signal field ($\Omega_s$). (b) A typical observed Rb
  Faraday resonance in which the transmission intensity for a cw
  signal field is shown as a function of magnetic field (i.e.,
  detuning of Zeeman levels from the two-photon resonance condition). The
full width of this resonance is 20 mG which
corresponds to a 15 kHz shift in the Zeeman levels.
  (c) Schematic of experimental setup.}
\end{figure}


\begin{figure}[ht]
\centerline{\epsfig{file=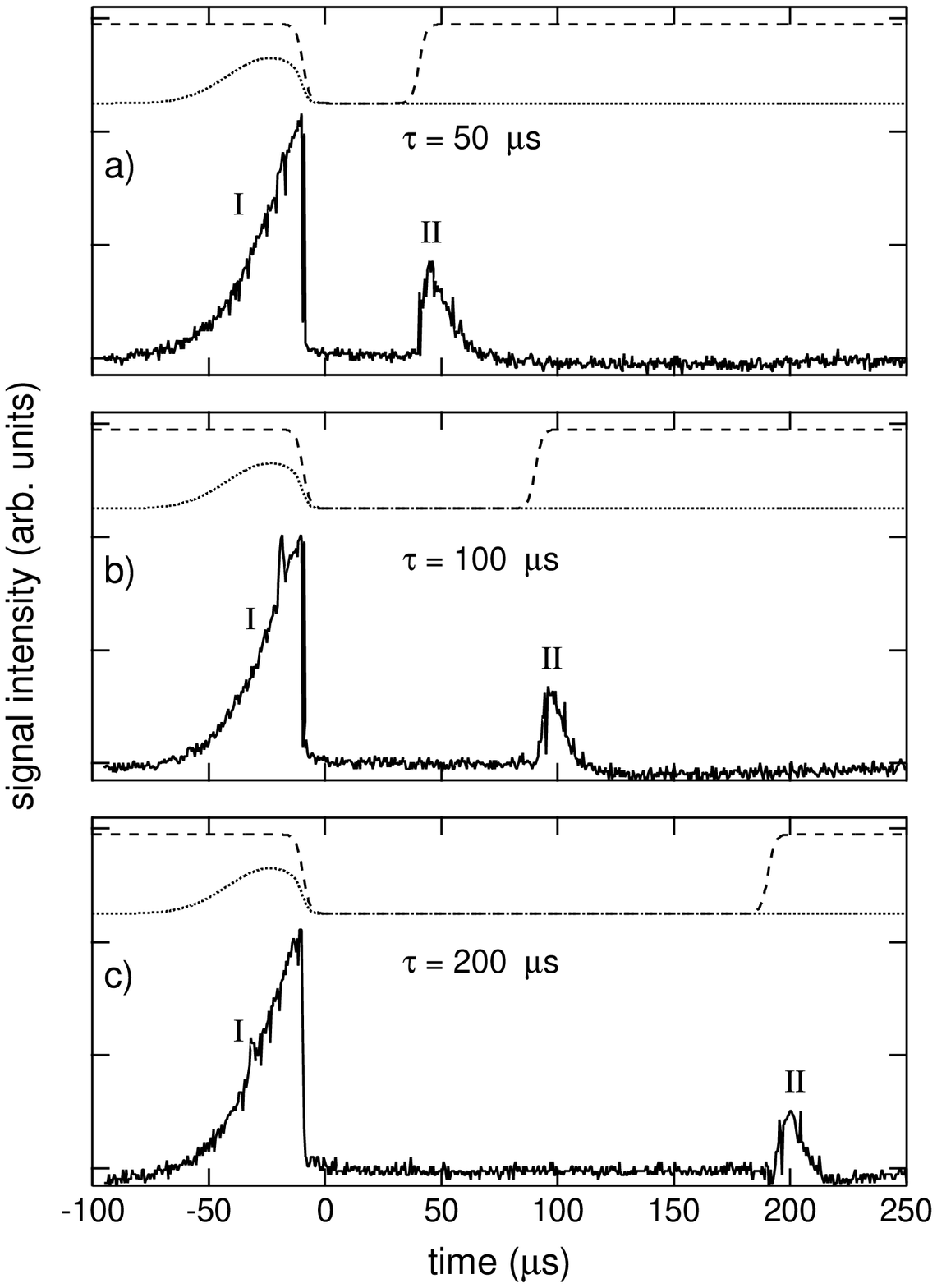,width=8.5cm}}
 \vspace*{2ex}
 \caption{Observed light pulse storage in a ${}^{87}$Rb vapor cell.
Examples are shown for storage times of (a) 50
   $\mu$s, (b) 100 $\mu$s, and (c) 200 $\mu$s. 
(Background transmission
 from the control field,
which leaks into the signal
   field detection optics, has been subtracted from these plots.)  Shown
above the data in each graph
are calculated representations of the applied control field (dashed
   line) and input signal pulse (dotted line).}
\end{figure}


\begin{figure}[ht]
\centerline{\epsfig{file=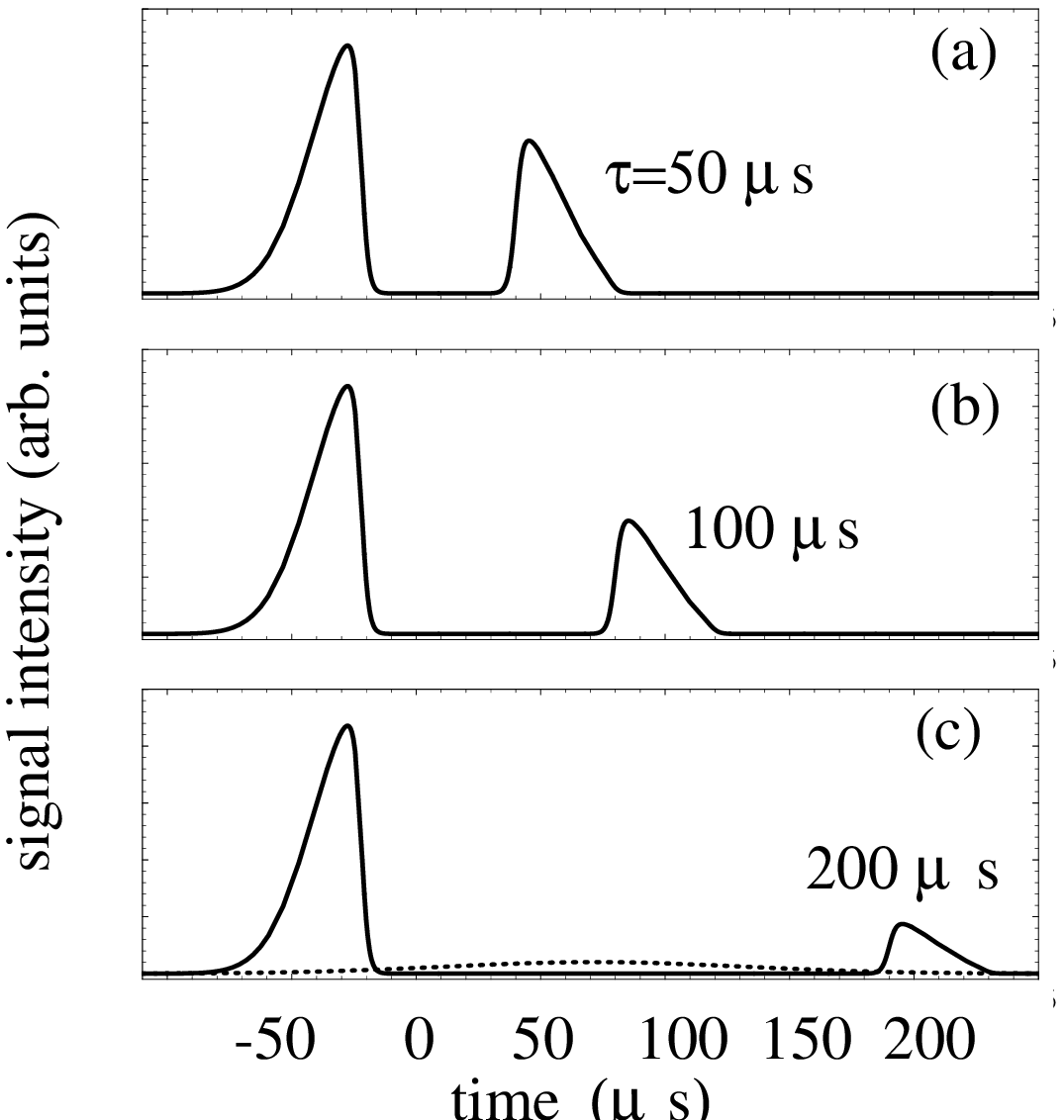,width=6.5cm}}
 \vspace*{2ex}
 \caption{Theoretical simulations of light storage in a three-state $\Lambda$
system. Solid curves in each graph correspond to the experimental
conditions of Fig. 2 (a-c).  Dephasing of the
coherence is modeled by an exponential with a decay constant of
150 $\mu$s. The dotted line in Fig. 3c corresponds to the usual EIT 
propagation with a cw control field of five
times weaker intensity than the input value of the dynamic case.}  

\end{figure}


\begin{thebibliography}{99}

\bibitem{stop}
M.~D.~Lukin, S.~F.~Yelin, and M.~Fleischhauer, Phys. Rev. Lett.
{\bf 84}, 4232 (2000); L.~M.~Duan, J.~I.~Cirac, and P.~Zoller (unpublished).

\bibitem{pol}
M.~Fleischhauer and M.~D.~Lukin, Phys.~Rev.~Lett. {\bf 84}, 5094 (2000);

\bibitem{slow} L.\ V.\ Hau, S.\ E.\ Harris, Z.\ Dutton, and C.\ H.\ Behroozi,
    Nature \textbf{397}, 594 (1999); M. Kash {\it et al.}
Phys. Rev. Lett. {\bf 82}, 5229 (1999); D.~Budker {\it et al.}
{\it ibid} {\bf 83}, 1767 (1999).

\bibitem{rev} See e.g. M.O.Scully and M.S. Zubairy, Quantum Optics 
(Cambridge University Press, Cambridge, UK, 1997);
S.~E.~Harris, Physics Today {\bf 50}, 36 (1997).


\bibitem{dis} Dissipative techniques for the partial transfer of quantum
statistics from light to atoms are reported in
A.~Kuzmich, K.~M\o lmer, and E.~S.~Polzik,
Phys. Rev. Lett. {\bf 79}, 4782 (1997) and
 J.~Hald, J.~L.~S\o rensen, C.~Schori, and E.~S.~Polzik,
Phys. Rev. Lett. {\bf 83}, 1319 (1999).

\bibitem{QED}  
J.~I.~Cirac, P.~Zoller, H.~Mabuchi, and H.~J.~Kimble,
Phys. Rev. Lett. {\bf 78}, 3221 (1997).

\bibitem{rempe} M.~Hennrich, T.~Legero, A.~Kuhn, and G.~Rempe, Phys. Rev.
Lett. {\bf 85}, 4872 (2000).

\bibitem{block} M.~D.~Lukin {\it et al.}, quant-ph/0011028.

\bibitem{tele} L.~Duan, J.~I.~Cirac, P.~Zoller and E.~Polzhik,
quant-ph/0003111.

\bibitem{squ} A.~Kuzmich, L.~Mandel, and N.~Bigelow,
Phys. Rev. Lett. {\bf 85}, 1594 (2000).

\bibitem{fre} O.~Kocharovskaya, Yu.~Rostovtsev, and M.~O.~Scully,
quant-ph/0001058.

\bibitem{nlo} H.~Schmidt and A.~Imamo\u glu, Opt. Lett. {\bf 21},
1936 (1996); S.\ E.\ Harris and Y.\ Yamamoto,
    \prl \textbf{81}, 3611 (1998);
S.~E.~Harris and L.~V.~Hau, {\it ibid} {\bf 82}, 4611 (1999);
for review see:
M.~D.~Lukin, P.~R.~Hemmer and M.~O.~Scully, in {\it Adv. At. Mol.
 and Opt. Physics} {\bf 42B}, 347 (Academic Press, San Diego, 2000).


\bibitem{lukin97} S.~E.~Harris, Phys. Rev. Lett. {\bf 70}, 552 (1993);
M.\ D.\ Lukin {\it et al.}, \prl
\textbf{79}, 2959 (1997).

\bibitem{hau} C.~Liu, Z.~Dutton, C.~H.~Behroozi, and L.~V.~Hau, submitted to 
Nature.

\end{thebibliography}
\end{document}